\documentclass[12pt]{article}
\usepackage{latexsym,epsfig,amssymb, amsmath, amsfonts, units, cite, mcite, todonotes}
\usepackage{enumerate}
              \newcommand{\rf}[1]{(\ref{#1})}

\def\bfone{\relax{\rm 1\kern-.35em 1}}

\newcommand{\be}{\begin{equation}}
\newcommand{\ee}{\end{equation}}
\newcommand{\ben}{\begin{displaymath}}
\newcommand{\een}{\end{displaymath}}
\newcommand{\bea}{\begin{eqnarray}}
\newcommand{\eea}{\end{eqnarray}}

\newcommand{\bean}{\begin{eqnarray*}}
\newcommand{\eean}{\end{eqnarray*}}

\newcommand{\vp}{\varphi}

\def\Kahler{K\"{a}hler~}

\def\K{K{\"a}hler}

\makeatletter \@addtoreset{equation}{section} \makeatother

\begin{document}

\begin{titlepage}

\begin{flushright}
\small ~ \\
\end{flushright}

\bigskip

\begin{center}

\vskip 2cm

{\LARGE \bf Superconformal Inflationary $\alpha$-Attractors} \\

\vskip 1.0cm

{\bf Renata Kallosh, Andrei Linde and Diederik Roest} \\

\vskip 0.5cm

{\em Department of Physics and SITP \\ 
Stanford University\\
 Stanford, California
94305 USA\\

{\small {{kallosh@stanford.edu  alinde@stanford.edu}}}} \\

\vskip 0.5cm

{\em Centre for Theoretical Physics \\
 University of Groningen \\
Nijenborgh 4, 9747 AG Groningen, The Netherlands\\
{\small {{d.roest@rug.nl}}}} \\

\end{center}

\vskip 2cm

\begin{center} {\bf ABSTRACT}\\[3ex]

\begin{minipage}{13cm}
\small

Recently a broad class of superconformal inflationary models was found leading to a universal observational prediction $n_s= 1-{2\over N} $ and $r={12\over N^2}$ \cite{Kallosh:2013hoa}. Here we generalize this class of models by introducing a parameter $\alpha$ inversely proportional to the curvature of the inflaton \Kahler manifold.  In the small curvature (large $\alpha$) limit, the observational predictions of this class of models coincide with the predictions of generic chaotic inflation models. However, for sufficiently large curvature (small  $\alpha$), the predictions converge to the universal attractor regime with $n_s= 1-{2\over N} $ and $r=\alpha {12\over N^2}$, which corresponds to the part of the $n_{s}- r$ plane favored by the Planck data. 

\end{minipage}

\end{center}

\vspace{2cm}

\vfill

\end{titlepage}

%%%%%%%%%%%%%%%%%%%%%%%%%%%%%%%%%%%%%%%%%%%%%%
%%%%%%%%%%%%%%%%%%%%%%%%%%%%%%%%%%%%%%%%%%%%%%

%\tableofcontents

\section{Introduction}

For a long time, it seemed rather difficult to realize generic chaotic inflation models  \cite{Linde:1983gd} in the context of supergravity. This problem was solved back in 2000 for the simplest theory $m^{2}\phi^{2}$ in \cite{Kawasaki:2000yn}. Since that time, we learned how to implement chaotic inflation with practically any desirable potential in  supergravity with chiral matter multiplets \cite{Kallosh:2010ug} as well as in supergravity with vector and tensor multiplets  \cite{Ferrara:2013rsa}. This allows to provide a consistent supergravity interpretation of any possible set of the observational parameters $n_{s}$ and $r$. Now that we have the freedom of choice, it is especially interesting to identify the models which could provide the most natural description of the available observational data.

In this paper we will continue our recent investigation \cite{Kallosh:2013hoa,Kallosh:2013pby,Kallosh:2013lkr,Ferrara:2013rsa,Kallosh:2013maa,Kallosh:2013tua} 
of a large family of different inflationary theories which lead to identical observational predictions in the limit of large number of e-folds $N$:
 \begin{align}\label{nsr}
  n_s  = 1-\frac{2}{N}\, , \qquad 
  r  = \frac{12 }{N^2} \,.
 \end{align}
These values correspond to the central part of the area in the $n_{s}-r$ plane favored by WMAP9 \cite{Hinshaw:2012aka} and Planck 2013 \cite{Ade:2013rta}. The models with $n_{s}$ and $r$ given by (\ref{nsr}) include the Starobinsky model $R+R^{2}$ \cite{Starobinsky:1980te}, the chaotic inflation model $\lambda \phi^{4}$ with non-minimal coupling to gravity ${\xi\over 2}\phi^{2} R$  \cite{Salopek:1988qh,Linde:2011nh,Kaiser:2013sna} for $\xi \gtrsim 0.1$, as well as a large set of various conformal, superconformal, and supergravity generalizations of these models, see \cite{Kallosh:2013pby,Kallosh:2013lkr,Kallosh:2013hoa,Ferrara:2013rsa,Kallosh:2013maa,Kallosh:2013tua,Ellis:2013xoa,Buchmuller:2013zfa}.

In particular, the prediction (\ref{nsr}) is a nearly universal feature of a very broad class of different models with spontaneously broken conformal or superconformal invariance found in \cite{Kallosh:2013hoa}. These models were generalized in \cite{Kallosh:2013maa} to the models with arbitrary {\it negative} non-minimal coupling to gravity ${\xi\over 2}\phi^{2} R$, $\xi < 0$. Predictions of this class of models continuously interpolate between the standard predictions of the chaotic inflation scenario with various potentials $V(\phi)$ and $\xi = 0$ \cite{Linde:1983gd}, and the attractor point (\ref{nsr}).

Recently we considered models with a large family of potentials $V(\phi)$ and introduced a generalized version of non-minimal coupling to gravity, such as  $\xi \sqrt{V(\phi)} R$, or $\xi\phi R$  \cite{Kallosh:2013tua}. For small $\xi$, these models have the same predictions as the usual inflation models with minimal coupling to gravity, but in the large $\xi$ limit their predictions converge to (\ref{nsr}). 

The inflaton potential of the canonically normalized inflaton field $\vp$ in all models yielding the universal result (\ref{nsr}) can be represented as 
\be
V(\vp) = V_{0}\left(1- e^{-\sqrt {2\over 3} \varphi}+...\right)
\ee 
in the limit $\vp \to \infty$. More general potentials $V_{0}(1- e^{-b \varphi}+...)$
have been considered in many inflationary theories, starting from \cite{Goncharov:1983mw}, with different values of the parameter $b$. However, in the context of the cosmological attractors discussed so far, the constant $b$ was always equal $\sqrt {2/3}$. In this paper we will consider two classes of supergravity models, where the potentials at large values of the inflaton field $\vp$ are given by 
\be
  V(\vp) = V_{0}\left(1- e^{-\sqrt {2\over 3\alpha} \varphi}+...\right) \,.
\ee 
A striking feature is that the new parameter $\alpha$ in both classes of models is related in the same way to the \K\, curvature $R_K$ of the inflaton's scalar manifold:
$
R_{K}=  -{2\over 3 \alpha }  \label{curvature}
$.
The scalar curvature thus that plays a crucial role in the generalized attractors that we put forward in this paper.

\begin{figure}[t!]
\centering
\includegraphics[scale=.5]{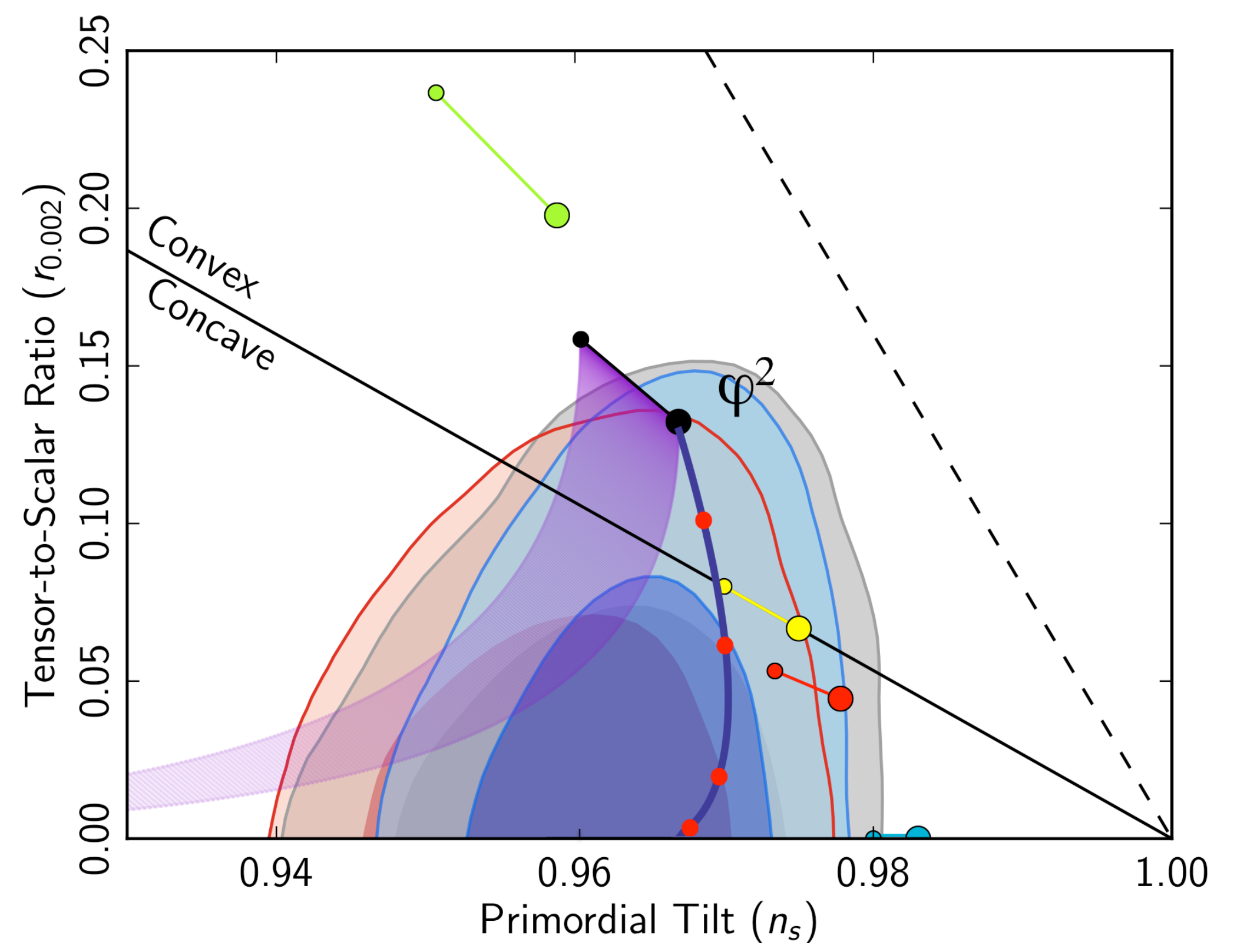} 
\caption{\it \small{The cosmological observables $n_s$ and $r$ for the theory with a potential $V_{0} \Bigl(1- e^{-\sqrt {2\over 3\alpha} \varphi}\Bigr)^2$ for $N=60$. As shown by the thick blue line, $n_s$ and $r$ for this model depend on $\alpha$ and continuously interpolate between the prediction of the simplest chaotic inflationary model with $V \sim \varphi^{2}$ for $\alpha \rightarrow \infty$, the prediction of the Starobinsky model for $\alpha = 1$ (the lowest red circle), and the prediction $n_{s} =1-\frac{2}{N}$, $r = 0$ for $\alpha \to 0$. The red dots on the thick blue line correspond to $\alpha = 10^{3},\, 10^{2},\, 10,\, 1$, from the top down.}}
\label{fig:interpol}
%\vspace{-.3cm}
\end{figure}

One of such models is an ${SU(1,1)\over U(1)}$ $\alpha$-$\beta$  model found in \cite{Ferrara:2013rsa} in a particular version of supergravity where the inflaton field is a part of a vector multiplet rather than a chiral multiplet.  The potential of the model is 
\be\label{alphanew}
V\sim \big(\beta- \alpha e^{-\sqrt {2\over 3\alpha} \varphi}\big)^2 \ .
\ee
The values of $n_s$  and $r$ for this model do not depend on $\beta$ and in the limit of large $N$ and small $\alpha$ are given by 
 \begin{align}
  n_s  = 1-\frac{2}{N}\,, \qquad 
  r  = \alpha \frac{12 }{N^2} \, .
\label{attr} \end{align}
 In Section \ref{ferrara} of this paper we analyze observational consequences of this model for generic values of $\alpha > 0$ and find that in the large $\alpha$ (small curvature) limit, the observational predictions for $n_s$ and $r$ of this class of models coincide with the predictions of the simplest chaotic inflation model with a quadratic potential, which are given by
\be\label{quadr}
n_{s} =1-{2\over N}\, ,\qquad  r = {8\over N} \ ,
\ee
for large $N$. The cosmological predictions of this theory continuously interpolate between \rf{attr} and \rf{quadr}, i.e. between the different $1/N$ universality classes identified in \cite{Mukhanov:2013tua, Roest:2013fha}. Note that the parameter $\alpha$ in this model can be much greater than 1, or it can be arbitrarily small, which leads to a very broad range of possible values of the tensor to scalar ratio $r$, see Figure \ref{fig:interpol} in Section 2.

The main aim of this paper is to find a generic set of models which have an attractor regime \rf{attr} at large $N$. In Sections 3 - 5 we will present a set of supergravities containing chiral superfields leading to this behavior, that we will refer to as $\alpha$-attractors. The new models generalize the class of superconformal inflationary models found in \cite{Kallosh:2013hoa}.  The potential of the inflaton field in the new class of models is an arbitrary function $f^2(\tanh {\varphi \over \sqrt {6 \alpha}})$. An interesting feature of the $\alpha$-attractors is that the relation between the geometric field $\Phi$ and the canonically defined field $\varphi$ reads
\be
  {\Phi \over \sqrt{3}} = \tanh {\varphi \over \sqrt {6 \alpha}} \,.
\label{rapidity}
\ee
This is fully analogous to the relation between velocity $v$ and rapidity $\theta$ in special relativity, ${v\over c} =  \tanh \theta$. The geometric non-canonical field ${|\Phi|} < \sqrt{3}$ has a limited range, analogous to velocity bound $v< c$;  in contrast, the rapidity and the canonical field $\vp$ have an unlimited range. 

The role of the new parameter $\alpha$, corresponding to the inverse curvature of the scalar manifold, is to regulate the relation between velocity and rapidity. When it increases above $\alpha=1$, the difference is diminished and there is a gradual transition from $\tanh{\varphi \over \sqrt {6 \alpha}}$ to $\vp \over \sqrt {6 \alpha}$. As this happens, $r$ grows and the predictions of the model move away from the `sweet spot' of the Planck data. In contrast, when $\alpha$ decreases below $\alpha=1$, the difference between rapidity and velocity becomes more pronounced. In this limit, $r$ moves below the value $\frac{12 }{N^2}$ \rf{nsr} and approaches the bottom of the Planck-favored region. As a result, predictions of the new set of models  continuously interpolate between the standard predictions of the chaotic inflation scenario with various potentials $V(\vp)$ \cite{Linde:1983gd}, and the universal attractor regime (\ref{attr}), see Figure 2 in Section 5.

The outline of this paper is as follows. In the next section we analyze the predictions of the $\alpha - \beta$ model. In Section 3 we review the broad class of $\alpha = 1$ superconformal models of \cite{Kallosh:2013hoa}. These models are generalized to arbitrary $\alpha$ in Section 4, and the resulting observational predictions are derived in Section 5. We end with our conclusions in Section 6.

\section{Phenomenology of the $\alpha-\beta$ model}\label{ferrara}

In this section we will study the $\alpha-\beta$ model with the potential (\ref{alphanew})  \cite{Ferrara:2013rsa}, which can be represented as
\be\label{alphanew1}
V= V_{0}\big(1- e^{-\sqrt {2\over 3\alpha} \varphi}\big)^2 
\ee
without loss of generality: this potential differs from (\ref{alphanew}) only by the overall normalization $V_{0}$ and by a shift of the field $\varphi$. This theory can be described in the context of supergravity with vector or tensor multiplets   \cite{Ferrara:2013rsa}, but, as we are going to show later, it can be also implemented in the theory with chiral multiplets. For  $\alpha = 1$, the potential \rf{alphanew1} coincides with the potential  of the Starobinsky model \cite{Starobinsky:1980te} in the scalar field representation \cite{Whitt:1984pd}. Meanwhile in the large $\alpha $ limit it coincides with the simplest chaotic inflation model with a quadratic potential. 
Indeed, for $\sqrt {2\over 3\alpha} \varphi \ll 1$, one has
\be\label{alphanew2}
V= V_{0}\big(1- e^{-\sqrt {2\over 3\alpha} \varphi}\big)^2 =  {m^{2}\over 2}\varphi^{2} \ .
\ee
where $m^{2} = {4V_{0}\over 3\alpha}$. 

This approximation is valid for $\varphi \ll \sqrt {3\alpha\over 2}$. Note that in the purely quadratic chaotic inflation model one has $N = \vp^{2}/4$. Therefore one can self-consistently describe inflation in the quadratic approximation \rf{alphanew2} for  $\alpha \gg 8N/3$, which yields the constraint $\alpha \gg 160$ for $N = 60$. In the large $N$ limit, in the quadratic approximation one has the same value of $n_{s} =1-\frac{2}{N}$ as in \rf{nsr}, but the value of $r$ is much higher, $r = {8\over N}$. 

By continuously decreasing $\alpha$ from $\infty$ to $0$, one can cover the full range of possible values of $r$ from $r = {8\over N}$ to $r = 0$. The last part of this trajectory, when $\alpha$ is of order one or smaller, proceeds along the attractor regime \eqref{attr}. The results of a numerical investigation of the parameters $n_{s}$ and $r$ in this model  are represented by a thick blue line in Figure 1.  

\section{Superconformal attractors for $\alpha = 1$}\label{alphaone}

In this section we will describe superconformal realizations of the $\alpha=1$ attractor models, following \cite{Kallosh:2013hoa}. In the next sections we will generalize these models for arbitrary $\alpha$ and study their observational consequences. 

We will consider 3 chiral supermultiplets: a conformon $X^0$,  an inflaton $X^1=\Phi$ and a sGoldstino $X^2=S$. The models are defined by two arbitrary functions of these superfields.  The first function is a \Kahler potential of the embedding manifold ${\cal N} (X, \bar X)$. It is real and has a Weyl weight 2. The second one represents a superpotential ${\cal W}(X)$. It is holomorphic and has a Weyl weight 3. The Lagrangian in terms of these functions is
 \begin{align}
  \mathcal{L} = \sqrt{-g} \left[ - \tfrac16 \mathcal{N}(X, \bar X) R - G_{I \bar J} \mathcal{D}^\mu X^I \mathcal{D}_\mu {\bar X}^{\bar J} - G^{I \bar J} \mathcal{W}_I \mathcal{\bar W}_{\bar J}
\right] \,,
 \end{align}
with $I, \bar I = \{ 0,1,2 \}$.

The models studied in \cite{Kallosh:2013hoa} have
\begin{equation}
\mathcal{N}\left( X,\bar{X}\right) =-\left| X^{0}\right|^{2}+\left|X^{1 }\right|^{2} +\left| S \right|^{2}.
\label{N}\end{equation} 
This \K\ potential  has a manifest $SU(1,1)$ symmetry between the  conformon and the inflaton, amongst others. The superconformal superpotential in \cite{Kallosh:2013hoa} was taken in the form
 \begin{align}
  \mathcal{W} = S  f(X^1 / X^0) \left[ (X^0)^2 - (X^1)^2 \right] \,.
 \end{align}
For $f=\rm{const}$ the superpotential preserves the $SO(1,1)$ subgroup of the $SU(1,1)$ symmetry of the \Kahler potential. However, the deviation of $f$ from a constant breaks this remaining symmetry.  The inflationary model, upon stabilization of the extra moduli, is given by 
 \begin{align}
  \mathcal{L} = \sqrt{-g} \left[ {1\over 2} R - {1\over 2}  (\partial \varphi)^2 - f^2\big (\tanh {\varphi\over \sqrt{6}} \big )
 \right] \, . 
   \end{align} 
For a very broad class of non-singular functions $f(z)$ with $z = X^1 / X^0$, which have a zero at some point $z$ and increase monotonically when $z$ grows up to $z=1$, one finds the universal prediction (\ref{nsr}) in the large $N$ limit. In particular, for the simplest set of functions $f(z) = \lambda z^{n}$, one has 
 \begin{align}
  V= f^2\big (\tanh {\varphi\over \sqrt{6}}\big )= \lambda^{2 }\tanh^{2n} {\varphi\over \sqrt{6}} \,.
 \end{align}
In this case expressions for $n_{s}$ and $r$ including higher order corrections in $1/N$ look as follows \cite{Roest:2013fha}:
 \begin{align}
  n_s & = 1-\frac{2}{N}+\frac{\sqrt{3  (4 n^2 + 3 )} - 3 n }{2 n N^2} +...\,, \qquad
  r = \frac{12 }{N^2}-\frac{6  \sqrt{3  (4 n^2 + 3 )}}{n N^3}+... \,.
 \end{align}
This underlines the attractor nature in the large-$N$ limit: all models have identical leading terms and only differ in subleading corrections.

\section{A general family of $\alpha$-attractors}

We now turn to the generalization of these superconformal models, leading to a family of $\alpha$-attractors. The superconformal \Kahler potential is now given by
 \begin{align}
  \mathcal{N}(X, \bar X)  = - |X^0|^2 \left[ 1 - \frac{|X^1|^2 + |S|^2}{|X^0|^2} \right]^\alpha \,.
 \end{align}
Note that the \Kahler potential only preserves the manifest $SU(1,1)$ symmetry between $X^0$ and $X^1$  for the special value $\alpha = 1$. The superconformal superpotential reads
 \begin{align}
  \mathcal{W} = S (X^0)^2 f(X^1 / X^0) \left[ 1 - \frac{(X^1)^2}{(X^0)^2} \right]^{(3 \alpha -1)/2} \,.
\label{calW} \end{align}
 The superpotential with a constant $f$ and $\alpha =1$ preserves the $SO(1,1)$ symmetry, the subgroup of $SU(1,1)$. However, when either $f$ is not constant, or $\alpha \neq 1$,  the $SO(1,1)$ symmetry is deformed.

In order to extract a Poincar{\'e} supergravity we gauge fix the conformal symmetry by setting $X^0 = {\bar X}^0 = \sqrt{3}$. The \Kahler and superpotential are then given by 
 \begin{align}
  K & = - 3 \alpha \log \left[1 - \frac{S \bar S+ \Phi \bar \Phi}{3} \right] \,, \qquad
  W = S f(\Phi / \sqrt{3}) (3 - \Phi^2)^{(3 \alpha -1)/2} \,.
\label{KW} \end{align}
Note that the conformal factor in the superpotential vanishes when $\alpha = 1/3$; this was exploited in a similar model with a specific choice of a linear function $f$ \cite{Scalisi}. 
For any real functions $f$, the   model above allows for a truncation to a one-field model via $S = \Phi - \bar \Phi = 0$; the stability of this truncation will be discussed below.
The  effective Lagrangian at $S=\Phi - \bar \Phi=0$ is
 \begin{align}
  \mathcal{L} = \sqrt{-g} \left[ {1\over 2} R - \frac{\alpha}{ (1- \Phi^2 / 3)^2}(\partial \Phi)^2 - f^2(\Phi / \sqrt{3})  \right] \, \,.
 \end{align}
Therefore the action is greatly simplified for real $\Phi$. As in  \cite{Kallosh:2013hoa}  we find a simple relation between the geometric field $\Phi$ and a canonical one $\varphi$: it is the rapidity-like relation given in \eqref{rapidity}.
The action for a canonical field  $ \varphi $ has an effective Lagrangian
 \begin{align}
  \mathcal{L} = \sqrt{-g} \left[ {1\over 2} R - {1\over 2}  (\partial \varphi)^2 - f^2\big(\tanh {\varphi\over\sqrt{6\alpha}}\big)
 \right] \, . 
 \label{action}  \end{align} 
At $f=$ const the potential of this model does not depend on $\vp$ and $\alpha$ and describes de Sitter vacuum.

In order to understand the role of the $\alpha$ parameter, we note that all $\alpha$-models during inflation at $S=0$ are defined by the $SU(1,1) / U(1)$  \K\, potential of the inflaton multiplet
\begin{align}
  K & = - 3 \alpha \log \left(1 - \frac{\Phi \bar \Phi}{3} \right) \,.   \end{align}
 This leads to kinetic terms of the form
\be
K_{\Phi\bar{\Phi}}\partial \Phi \partial \bar \Phi={\alpha\over  \bigl(1 - \frac{\Phi \bar \Phi}{3} \bigr)^2}  \partial \Phi \partial \bar \Phi \ .
\ee
This \K\, metric $g_{\Phi\bar \Phi}= K_{\Phi\bar{\Phi}}$ corresponds to an $SU(1,1)/U(1)$ symmetric space with the constant curvature:
 \be
R_{K}= -g^{-1}_{\Phi\bar \Phi}\partial_\Phi\partial_{\bar \Phi} \log g_{\Phi\bar \Phi} \,,
 \ee
given by \eqref{curvature}.
The same relation was found in the context of the supersymmetric $\alpha$-$\beta$ model in \cite{Ferrara:2013rsa}. Here we notice that
 all  $\alpha$-attractors of this paper with an arbitrary function $f\big (\tanh{\varphi\over \sqrt{6\alpha}}\big)$
 have a universal interpretation of the parameter $\alpha$,
 \be
\alpha =  -{2\over 3R_{K}}  \,,
 \ee
in terms of the $SU(1,1)/U(1)$ symmetric space with the negative constant curvature $R_K$ in this class of models.

Finally, we address the stability of the truncation to the single-field model. To this end we add a stabilisation term to our original superconformal \Kahler potential,
 \begin{align}
  \mathcal{N}(X, \bar X)  = - |X^0|^2 \left[ 1 - \frac{|X^1|^2 + |S|^2}{|X^0|^2} + 3 g\frac{|S|^4}{|X^0|^2 ( |X^0|^2 - |X^1|^2)} \right]^\alpha \,.
 \end{align}
The original four scalar fields have the following masses at the inflationary trajectory $S = \Phi - \bar \Phi = 0$:
 \begin{align}
  m_{{\rm Re}(\Phi)}^2 & = \eta_\varphi V \,, \quad
  m_{{\rm Im}(\Phi)}^2  = \left(2 - \frac{2}{3 \alpha} + 2 \epsilon_\varphi - \eta_\varphi \right) V \,, \quad
  m_S^2 = \left( \frac{12 g -2}{3 \alpha} + \epsilon_\varphi \right) V \,,
 \end{align}
where $\epsilon_\vp$ and $\eta_\vp$ are the slow-roll parameters of the effective single-field model \eqref{action}. In order to achieve stability up to slow-roll suppressed corrections, the second equation requires $\alpha > 1/3$ for stabilisation of the inflationary trajectory, and the latter requires $g > 1/6$. 

\section{Phenomenology of the $\alpha$-attractors}

In Sections 3 and 4 we studied superconformal $\alpha$-models. For the cases where $f(\Phi/3)$ is a holomorphic function, and the inflationary trajectory $S = \Phi - \bar \Phi = 0$ is stable, investigation of inflation is reduced to the study of a theory describing a single canonically normalized field $\vp$ with the Lagrangian  \rf{action}.

From a phenomenological point of view, one may also consider the purely bosonic theory \rf{action} on its own ground, and study its implications for a certain choice of functions and parameters $\alpha$. We may subsequently restrict ourself to the choice of the functions and parameters consistent with the superconformal origin of the theory.

The simplest class of models \rf{action} has a  potential involving a monomial  
 \begin{align}
  V=\tanh^{2n} (\varphi/\sqrt{6\alpha}) \,. \label{monomial-potential}
 \end{align}
These give rise to inflation for all values of $\alpha$; the nature of this process will change, however, as $\alpha$ is decreased. In particular, in the slow roll approximation the cosmological observables are given by the following expressions 
\begin{align}\label{exact1}
r (\alpha, n, N) & = {12  \alpha\over N^2+ {N\over 2n} g(\alpha,n) + {3\over 4}\alpha} \, ,  \\
\notag \\
n_s(\alpha, n, N) & = {1 - {2 \over N}   - {3 \alpha \over 4 N^2} +{ 1\over 2nN}(1 - {1\over N})g(\alpha,n)
\over 1+ {1\over 2nN} g(\alpha,n) + {3 \alpha\over 4N^2}} \, , \label{exact}
\end{align}
where
\be
g(\alpha,n)\equiv \sqrt{3 \alpha (4 n^2 + 3 \alpha)} \ .
\ee
These expressions are exact modulo the assumption 
 that inflation ends when $\epsilon = 1$. One can check that this violation of the slow-roll conditions occurs first, for all models with $\alpha > 1/3$ (provided $n \geq 1/2$) which is what we will concentrate on for the moment. This restriction on $\alpha$ follows from the stability analysis of the previous section, and we will find that it coincides with the observationally most interesting region.

We will first analyze the above expression for large $\alpha$. In the  limit $\alpha \to \infty$ one has 
 \begin{align}
  n_s & = 1-\frac{2n+2}{2N+n}\,, \qquad 
  r = \frac{16 n}{2N+n} \, ,
 \end{align}
 which coincide with the corresponding expressions for the theory $V(\vp) \sim \vp^{2n}$. We therefore recover from all models of the type $\tanh^{2n} (\varphi/\sqrt{6\alpha})$ the corresponding chaotic monomial models $\vp^{2n}$. This was expected from the relation between velocity and rapidity, as explained near \rf{rapidity}.   One can see there that at large $\alpha$ rapidity coincides with velocity. 
 
\begin{figure}[t!h!]
\centering
\includegraphics[scale=.5]{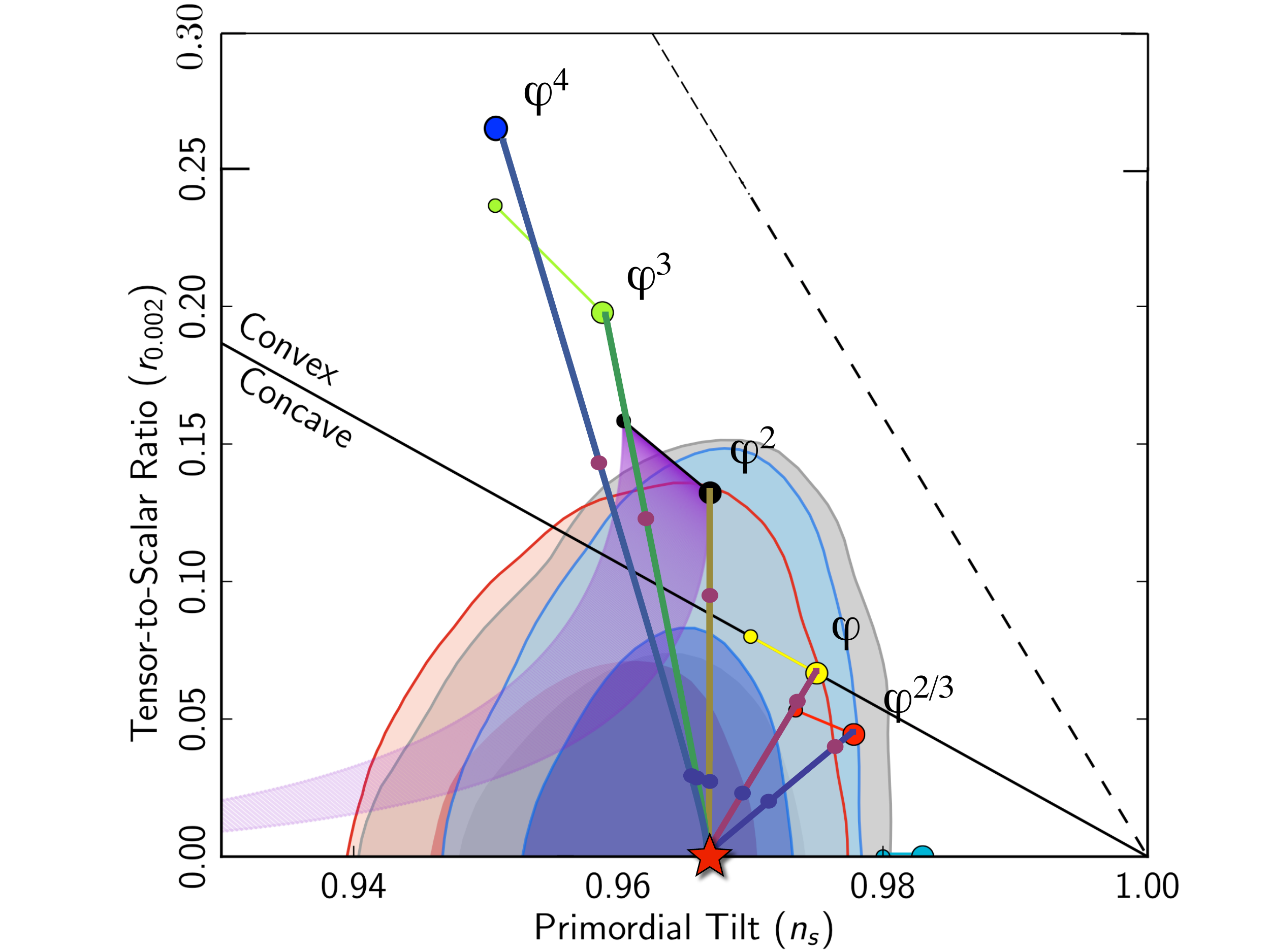} 
\caption{\it \small{ The cosmological observables $(n_s,r)$  for different scalar potentials $\tanh^{2n} ({\vp \over \sqrt{6 \alpha}})$ with $2n = (2/3, 1, 2, 3, 4)$  for $N=60$. These continuously interpolate between the predictions of the simplest inflationary models with the monomial potentials $\varphi^{2n}$ for $\alpha \rightarrow \infty$, and the attractor point $n_{s} =1-2/N$, $r = 0$ for $\alpha \to 0$, shown by the red star. The different trajectories  form a fan-like structure for $\alpha \gg n^2$. The set of dark red dots at the upper parts of the interpolating straight lines corresponds to $\alpha = 100$. The set of dark blue dots corresponds to $\alpha = 10$. The lines gradually merge for $\alpha = O(1)$.}}
\label{fig:ClassI}
%\vspace{-.3cm}
\end{figure}

  Secondly, for all $\alpha \gg n^{2}$ we find
 \begin{align}
  n_s & \approx 1-\frac{2}{N}-\frac{n-1}{8n} r\,, \qquad
  r \approx {24n\alpha\over N(2nN +3\alpha)} \,,
 \end{align}
where we have moreover used that $n \ll N$. Note that for $n^{2}\ll \alpha \ll nN$, $r$ grows linearly with the growth of $\alpha$, until it reaches its limiting value $r  \approx {8n\over N}$. Meanwhile the parameter $n_{s}$ linearly depends on $r$   for all $\alpha \gg n^{2}$. This gives rise to the  interpolation behavior as is shown in Figure \ref {fig:ClassI}: the linear trajectories directly interpolate between the chaotic inflation models $\vp^{2n}$ at large $\alpha$ and the universal attractor at small $\alpha$.

We now turn to the small $\alpha$ behavior close to the attractor, where $\alpha$ is of order one. When expanding \eqref{exact} in the large-$N$ limit we find
 \begin{align}\label{attractor}
  n_s & \approx 1-\frac{2}{N}+\frac{\sqrt{3 \alpha (4 n^2 + 3 \alpha)} - 3 n \alpha}{2 n N^2} \,, \qquad
  r \approx \frac{12 \alpha}{N^2}-\frac{6 \alpha \sqrt{3 \alpha (4 n^2 + 3 \alpha)}}{n N^3} \,.
 \end{align}
Note that  the supersymmetric models \rf{KW} are stabilized at  $\alpha>1/3$. We will therefore look at  the range of the attractor points near $\alpha=1$ where the value of $r$ increases by the order of magnitude:
\be
1/3 <\alpha <3 \,, \qquad         10^{-3} < r < 10^{-2} \ .
\ee
In this range one finds that the dependence on $n$ is very small: all $V=\tanh^{2n} (\varphi/\sqrt{6\alpha})$ models give approximately the same value of $r(\alpha)\approx \frac{12 \alpha}{N^2}$ and an $n$-independent value of $n_s$ in the leading large-$N$ approximation.  Therefore, according to these models, the expected level of gravity waves is flexible and uniquely defined by the curvature of the \K\, manifold $R_{K}=  -{2\over 3 \alpha }$ for this range of parameters. However, this behavior occurs only in a tiny part of the $n_{s}-r$ plane, covered by the red star. Soon after $\alpha$ exceeds $3$, the second non-universal term in \rf{attractor} becomes important and the trajectories separate as in a fan, covering a significant part of the dark blue area in the $n_{s}-r$ plane of Figure \ref{fig:ClassI}.

Our numerical investigation was done for the simplest choice of the function $f(\Phi/ \sqrt{3}) \sim \Phi^{n}$. If one considers a more complicated function, e.g.~$f(\Phi/ \sqrt{3}) = \bigl({\Phi\over\Phi+ \sqrt{3}}\bigr)^{n}$, one finds a different family of trajectories interpolating between various models of chaotic inflation and the attractor regime (\ref{attr}). In particular, for $n = 1$ one recovers the potential  \rf{alphanew1}, and the interpolating regime shown in Figure \ref{fig:interpol}, Section 2.

\section{Discussion}

Our results represent the generalization of the attractor values \eqref{nsr} for the inflationary parameters $n_{s}$ and $r$, which have appeared in a variety of contexts, to the family of attractor values \eqref{attr} labelled by the parameter $\alpha$. In Section 2, we have developed the phenomenology of the $\alpha - \beta$ model \cite{Ferrara:2013rsa}. In Sections 3 - 5 we 
proposed  a new class of $\alpha$-attractors, generalizing the attractors found in \cite{Kallosh:2013hoa}. The parameter $\alpha$ has the same geometrical interpretation in both types of models, corresponding to the inverse curvature of the underlying scalar manifold of the inflaton's supermultiplet, $R_{K} = -{2\over 3\alpha}$. Its role in the present discussion is to control the distinction \eqref{rapidity} between the geometric fields ${\Phi} = \sqrt{3}\tanh {\varphi \over \sqrt {6 \alpha}}$ and canonical fields $\vp$, akin to that between velocity and rapidity in special relativity.

The potential in the new class of models is  an arbitrary function of ${\Phi} = \sqrt{3}\tanh {\varphi \over \sqrt {6 \alpha}}$.
As a particular example, we studied simplest versions of such models with $V(\Phi) \sim \Phi^{n}$. For large $\alpha$, these models yield the same observational predictions as the conventional chaotic inflation with $V(\vp) \sim \vp^{n}$. Therefore this family of models provide a continuous interpolation between the chaotic inflation values for $(n_s,r)$ and the universal attractor values \eqref{attr}, as follows from \rf{exact1}, \rf{exact}. 

A number of regimes is of particular interest:
 \begin{itemize}
 \item
In the limit $\alpha \to \infty$ ($R_{K} \to 0$, flat \K\ manifold), this set of models reproduces predictions of various chaotic inflation models. However, when $\alpha$ decreases towards $\alpha = O(1)$ ($|R_{K}|$ grows to $|R_{K}| =O(1)$), these predictions start converging to the universal attractor point, as shown in the fan-like attractor in Figure 2. 
\item
For $\alpha=1$ and  $R_{K}=  -{2\over 3  }$ we have a special case when the underlying \K\, potential of the embedding manifold has an unbroken $SU(1,1)$ symmetry and the superpotential has an $SO(1,1)$ subgroup of this symmetry deformed only by the deviation of the function $f(\tanh {\vp \over \sqrt {6}})$ from a constant value \cite{Kallosh:2013hoa}. 
\item 
For moderate deviations from this special value, of the order $1/3 < \alpha < 3$, the tensor to scalar ratio $r$ varies universally between $10^{-3} < r < 10^{-2}$. This entire regime, with $-2 < R_K < - {2 \over 9}$, is therefore subject to the attractor behavior \rf{attr}.
 \item
 If we instead take $\alpha$ smaller than $1/3$, in the context of the present superconformal realization of the $\alpha$-attractors with chiral superfields, instead of the vector one, the scalar superpartner of the inflaton has a negative mass squared: the model becomes unstable in this limit with $\alpha < 1/3$. However,
 the bosonic model \rf{action}, as well as the $\alpha - \beta$ model \rf{alphanew} \cite{Ferrara:2013rsa}, do not have stability issues even for very small $\alpha$ and they display the attractor behavior with the universal values approaching $r=0$ as $\alpha \rightarrow 0$.  \end{itemize}
 
The existence of the universal attractor regime in a large set of different inflationary models, converging in the same area of the $(n_{s},r)$ plane, is an amazing fact requiring further investigation. It does not necessarily mean that the evolution of our part of the universe must be described by one of such theories, but it certainly provides a very interesting target for future theoretical and observational explorations. 

\section*{Acknowledgements}

We acknowledge stimulating discussions with R. Bond and with our collaborators on a related project, S. Ferrara and M. Porrati.  RK and AL are supported by the SITP and by the NSF Grant PHY-1316699 and RK is also supported by the Templeton foundation grant `Quantum Gravity Frontiers'. DR would like to thank the SITP for its warm hospitality and NWO for financial support with a VIDI grant.

\end{document}